# Incident angle dependence of reactions between graphene and hydrogen atom by molecular dynamics simulation

Seiki Saito[a], Atsushi Ito[b], and Hiroaki Nakamura[a,b]

[a] *Department of Energy Engineering and Science, Graduate school of Engineering, Nagoya University, Furo-cho, Chikusa-ku, Nagoya 464-8603, Japan*

[b] *National Institute for Fusion Science, 322-6 Oroshi-cho, Toki 509-5292, Japan*

Incident angle dependence of reactions between graphene and hydrogen atoms are obtained qualitatively by classical molecular dynamics simulation under the *NVE* condition with modified Brenner reactive empirical bond order (REBO) potential. Chemical reaction depends on two parameters, i.e., polar angle $\theta$ and azimuthal angle $\phi$ of the incident hydrogen. From the simulation results, it is found that the reaction rates strongly depend on polar angle $\theta$. Reflection rate becomes larger with increasing $\theta$, and the $\theta$ dependence of adsorption rate is also found. The $\theta$ dependence is caused by three dimensional structure of the small potential barrier which covers adsorption sites. $\phi$ dependence of penetration rate is also found for large $\theta$.

## 1. Introduction

Carbon fiber composites (CFC) is regarded as one of candidates of the divertor plate of nuclear fusion reactor and it is considered to be bombarded with hydrogen plasma. The hydrogen plasma erodes the divertor plate, yielding $H_2$ and other hydrocarbon molecules such as $CH_x$ and $C_2H_x$, which are undesirable impurities in plasma confinement experiments. To understand the mechanism of chemical and physical interactions between hydrogen plasma and the divertor plate, it is necessary to clarify the elementary processes of the reactions. Molecular dynamics (MD) simulation, where the equation of motion of the atoms is solved numerically, is a powerful tool to research the elementary processes. In the previous research, we showed that the reaction depends strongly on the incident energy of hydrogen atom. We also clarified its physical mechanism in the case of perpendicular injection against a graphene sheet by MD simulation [1].

The surface of the divertor plate is not flat in the nanoscale level any longer. For example, CFC has such a structure that several thousand carbon fibers are twisted. The fibers have polycrystalline structure with clusters measuring approximately 1 nm to 1 μm, and the crystal axes of the clusters are not aligned. Moreover, single crystalline graphite has the structure that a lot of graphene sheets are layered. In this case, oblique incidence as well as vertical incidence should be taken into account to understand the reaction of the diverter plate. In the present paper, we investigate the incident angle dependence of the reactions between a hydrogen atom and a graphene sheet by MD simulation.

In the following sections, simulation method will be described in section 2, and then, the energy dependence of the reaction rates for vertical and oblique incidences will be shown in section 3, followed by the summary given in section 4.

## 2. Simulation Method

### 2.1 Simulation algorithm

In the present research, we used a classical molecular dynamics simulation under the *NVE* condition with modified Brenner reactive empirical bond order (REBO) potential [2, 3]. REBO potential is one of widely popular potential for MD simulation of carbon systems. A carbon atom has four or less covalent bonds which are derived by its four valence electrons. The type of bonding depends strongly on the bonding state of carbon atom. REBO potential gives us the sufficient information of behavior of atoms by the following form:

$$U \equiv \sum_{i,j>i}\left[V^{R}_{[ij]}(r_{ij}) - \bar{b}_{ij}(\{r\},\{\theta^{B}\},\{\theta^{DH}\})V^{A}_{[ij]}(r_{ij})\right], \quad (1)$$

where $r_{ij}$ is the distance between the *i*-th and *j*-th atoms. The function $V^{R}_{[ij]}$ and $V^{A}_{[ij]}$ represent repulsion and attraction, respectively. The $\bar{b}_{ij}$ generates a many-body force. Second-order symplectic integration is used to solve the time evolution of the equation of motion [4] with the time step $5\times10^{-18}$ s.

*author's e-mail: saito.seiki@nifs.ac.jp*

## 2.2. Simulation model

As shown in Fig. 1(a), one hydrogen atom is injected to a graphene sheet. The graphene sheet consists of 160 carbon atoms with periodic boundary condition in the $x$ and $y$ directions. The initial temperature of the carbons is set to zero Kelvin.

Figure 1(b) shows the $O$-$xyz$ coordinate of our simulations with polar angle $\theta$ and azimuthal angle $\phi$. The hydrogen atoms are injected into the graphene from $z = 3$ Å. The $x$ and $y$ coordinates of the injection position of the hydrogen are set randomly. The incident angle is chosen ($\theta$, $\phi$). In the second simulation, we set the simulation model to the same one except the $x$ and $y$ coordinates of injection position as the initial state of the first simulation. We change the $x$ and $y$ coordinates of injection point randomly and start the second simulation. We repeat the above simulation 2500 times for each ($\theta$, $\phi$) individually. We observed three types of reactions, i.e., reflection, adsorption, and penetration.

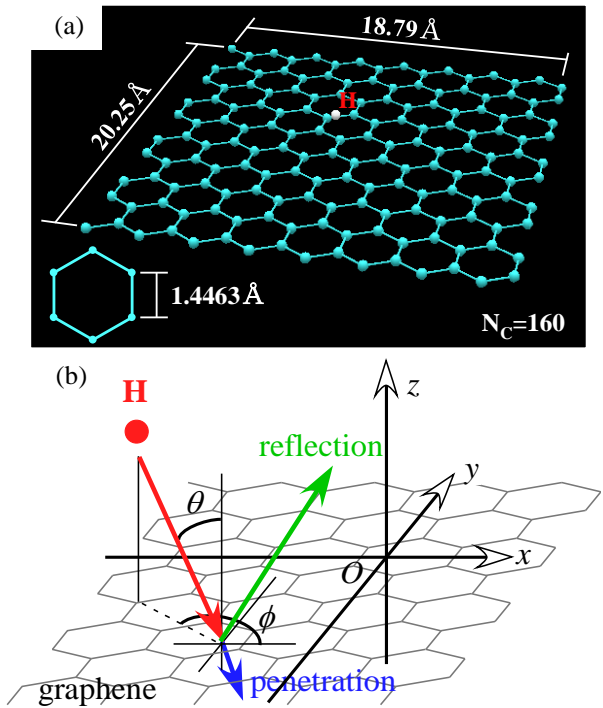

Fig. 1 (a) Simulation model and (b) the $O$-$xyz$ coordinate of our simulations with polar angle $\theta$ and azimuthal angle $\phi$. The center of mass of the graphene is set to the origin of coordinate.

## 3. Simulation Results

### 3.1 Energy dependence of reaction rate for vertical incidence

In this section, the energy dependence of vertical injection of hydrogen will be shown to compare with the dependence for oblique injection. Figure 2 shows incident energy dependence of the adsorption, reflection, and penetration rates for $\theta = 0°$. Different interactions are dominant for different incident energies $E_{in}$.

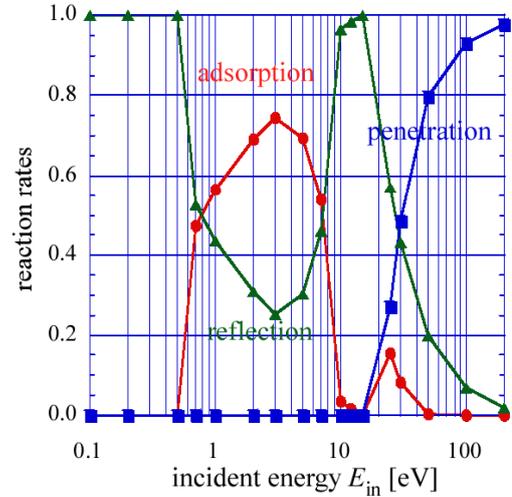

Fig. 2 Incident energy dependence of the adsorption, reflection, and penetration rates for vertical incidence.

In the case of the low incident energy, i.e., $E_{in} < 0.5$ eV, almost all injection atoms are reflected. To understand the physical mechanism, we draw a potential energy contour plot in the $y$-$z$ plane which includes two adjoined carbon atoms (Fig. 3). The two adjoined carbon atoms are located at $(y, z) = (1.4463 \text{Å}, 0 \text{Å})$ and $(-1.4463 \text{Å}, 0 \text{Å})$. There is an adsorption site 1.1 Å above each of the carbons. We found that there is a dome-shaped small potential barrier, the strength of, which is about 0.5 eV, in front of adsorption sites. Adsorbed hydrogen atoms must have larger perpendicular component of kinetic energy to the tangential plane of the dome-shaped barrier than the potential barrier. The three dimensional structure of this barrier also affects the adsorption for oblique injection. We will discuss this point in detail in section 3.2. For $0.5 \text{ eV} < E_{in} < 7 \text{ eV}$, adsorption becomes dominant, but reflection becomes dominant again for $7 \text{ eV} < E_{in} < 30 \text{ eV}$. To be adsorbed to the adsorption site, hydrogen has not only to have enough kinetic energy to pass the small potential barrier but also to lose its kinetic energy so that hydrogen may fall into the adsorption site by giving its kinetic energy to the surrounding carbons. Hydrogen atoms injected just on carbon atoms do not have enough reaction time to lose its kinetic energy to be trapped. Increasing incident energy makes the reaction time shorter, and more incident hydrogen atoms are reflected.

Penetration is the dominant process for $E_{in} > 30$ eV. In the case of $E_{in} = 25$ eV, the second peak of adsorption is observed. This peak is caused by the back side adsorption. Penetrating the graphene sheet, the injected hydrogen atoms lose approximately 25 eV of its kinetic energy so that it can be adsorbed at the back side of graphene sheet.

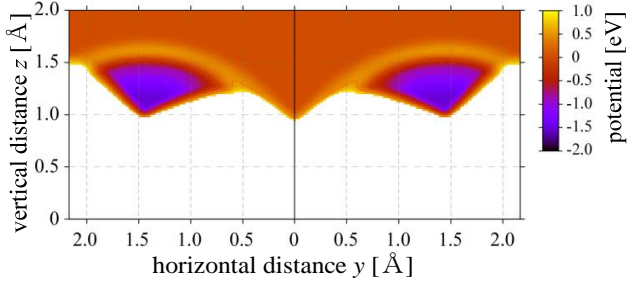

Fig. 3 Potential energy contour plot in the *y-z* plane which includes two adjoined carbon atoms.

## 3.2 Oblique injection – $\theta$ dependence

Figure 4 shows the incident energy dependence of reaction rates for $\phi=0°$ on the angle $\theta$. The reflection rate becomes large with increasing $\theta$ (Fig. 4(a)). The rising point of penetration rate shifts to high energy side with increasing $\theta$ (Fig. 4(b1)). This shift is caused by the fact that the vertical component of incident energy decreases with $\cos^2\theta$. As shown in Fig. 4(b2), the penetration rates are described as a function of $E_{in}\cos^2\theta$. On the contrary, the adsorption rate can not be written as a function of $E_{in}\cos^2\theta$ (Fig. 4(c)). The difference between penetration and adsorption rates comes from that the adsorption rate depends on the three dimensional structure of potential surface on graphene sheet. That is reason why the adsorption rate has more complicated $\theta$ dependence than the penetration rate.

Figure 5 shows reaction maps in the case of $E_{in}=0.5$eV, $\phi=0°$. Red dots imply that the hydrogen atom is adsorbed. Green and blue dots denote reflection and penetration, respectively, in Figs. 5 and 7(b). We plot, in Fig. 5 "impact points" without interactions between carbon and hydrogen, which are defined by

$$\begin{pmatrix} x^i \\ y^i \\ z^i \end{pmatrix} \equiv \begin{pmatrix} x_0^i \\ y_0^i \\ z_0 \end{pmatrix} - \frac{z_0}{v_{0z}^i}\begin{pmatrix} v_{0x}^i \\ v_{0y}^i \\ v_{0z}^i \end{pmatrix}, \qquad (2)$$

where $^t(x_0^i, y_0^i, z_0)$ is the initial location and $^t(v_{0x}^i, v_{0y}^i, v_{0z}^i)$ is the initial velocity of the *i*-th injected hydrogen atom. Moreover, $z_0$ is equal to 3 Å for all injections. As shown in Fig. 5, some of injected hydrogen atoms are adsorbed for $E_{in}=0.5$ eV and $\theta=20°$ although all injections are reflected for vertical incidence. The reason is that the potential barriers around $y=\pm 0.8$ Å have the minimum value 0.45 eV. That is, not only the fact that the injected hydrogen atom comes to the minimum point of potential barrier but also vertically injection into tangential plane of the dome-shaped barrier stands up in front of injected hydrogen. In the case of incident energy $E_{in}=5$ eV, adsorption rate for $\theta=80°$ is much lower than low $\theta$ case even though the incident energy is 10 times higher than the barrier. The small potential barrier behaves a dominant role for high incident energy in the case that polar angle $\theta$ is large.

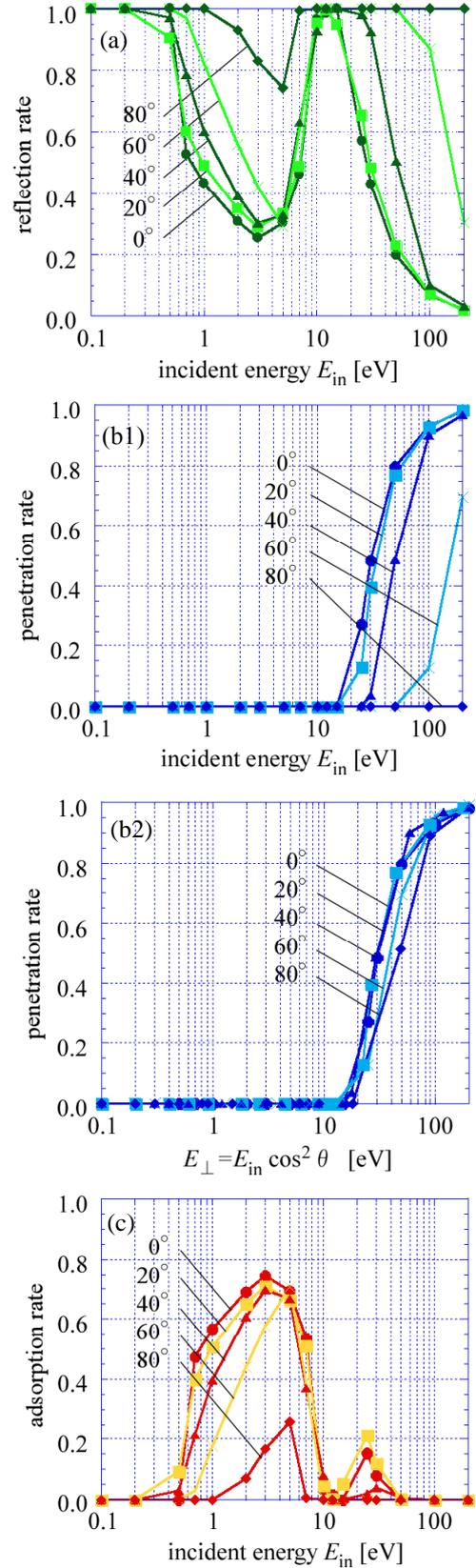

Fig. 4 Incident energy dependence of reflection(a), penetration(b1), and adsorption rates(c) with different $\theta$ in the case of oblique incidence. The graph (b2) shows the penetration rate against vertical component of incident energy $E_{in}\cos^2\theta$.

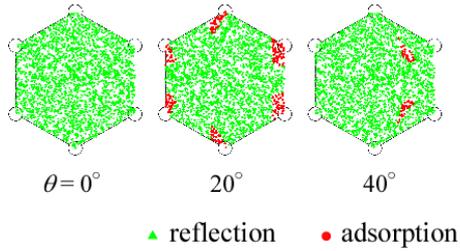

▲ reflection    ● adsorption

Fig. 5 Reaction maps for $E_{in} = 0.5$ eV, $\phi = 0°$.

### 3.3. Oblique injection - $\phi$ dependence

Figure 6 shows the $\theta$ dependence of reaction rates for $E_{in} = 5$ eV with a several azimuthal angles $\phi$. This figure shows that the reaction rates do not change by $\phi$, especially for adsorption rate. This property is caused by the fact that the reaction of adsorption depends on the structure of the small potential barrier as mentioned above, and the barrier has $\phi$ symmetry, that is, 6-fold rotational symmetry of graphene sheet.

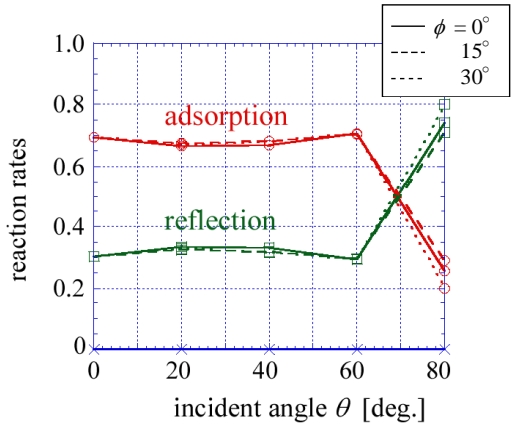

Fig. 6 The $\theta$ dependence of reaction rates for $E_{in} = 5$ eV with each $\phi$.

We found that there is $\phi$ dependence of penetration rate for $E_{in} = 100$ eV and $\theta \geq 60°$ as shown in Fig. 7(a). In such high incident energy case, penetration is dominant process, and the small potential barrier is regarded to be negligible. Thus, it is considered that the penetration process is simple mechanism. The injection hydrogen, which collides with carbon atoms, are reflected, and the other injection hydrogen atoms penetrate through a graphene sheet. The penetration rate becomes smaller with increasing $\theta$, because the shadow area, which is defined as the area where hydrogen is interrupted to inject onto the graphene surface by carbons, becomes larger with increasing $\theta$. In the case of $\phi = 30°$, the shadow reaches to the next carbon atom and all injection atoms that are connected to two carbon atoms in diagonal are reflected for $\theta = 80°$ as shown in Fig. 7(b). This overlapping condition of the shadow would be changed for a different value of $\phi$. This is the reason why the penetration rate has $\phi$ dependence when $\theta$ is large.

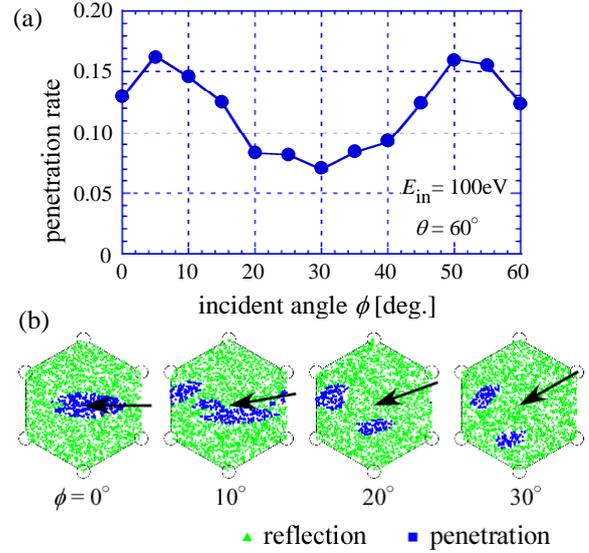

▲ reflection    ■ penetration

Fig. 7 (a) $\phi$ dependence of penetration rate and (b) the reaction maps for $E_{in} = 100$ eV, $\theta = 60°$.

### 4. Summary

Incident angle dependence of the adsorption, reflection, and penetration rates are calculated by MD simulation. As hydrogen atoms are injected along the horizontal direction against graphene sheet, the reflection becomes larger. The penetration rate is considered to be a function of the vertical component of incident energy $E_{in}\cos^2\theta$. The adsorption rate has more complicated $\theta$ dependence than the penetration rate. This complexity is derived by the three dimensional structure of the dome shaped small potential barrier, the height of which is about 0.5eV. It was also found that the adsorption rate does not depend on $\phi$. The penetration rate for $\theta \geq 60°$ has $\phi$ dependence.

### Acknowledgment

Numerical simulations were carried out by use of the Plasma Simulator at National Institute for Fusion Science. The work is supported by the National Institutes of Natural Sciences undertaking Forming Bases for Interdisciplinary and International Research through Cooperation Across Fields of Study and Collaborative Research Program (No. NIFS09KEIN0091) and Grants-in-Aid for Scientific Research (No. 19055005) from the Ministry of Education, Culture, Sports, Science and Technology.